\documentclass[letterpaper,twocolumn,10pt]{article}
\usepackage{usenix-2020-09}

\usepackage{tikz}
\usetikzlibrary{positioning}
\usepackage{amsmath}
\usepackage{xspace}
\usepackage{booktabs}
\usepackage{multirow}
\usepackage{enumitem}
\usepackage{listings}
\usepackage{comment}

\newcommand{\sys}{Patchwork\xspace}
\newcommand{\todo}[1]{\textcolor{red}{[TODO: #1]}}
\newcommand{\hzx}[1]{\textcolor{blue}{[hzx: #1]}}

\graphicspath{{figures/}}

\begin{document}
%-------------------------------------------------------------------------------

%don't want date printed
\date{}

\title{Replicate What Varies: Resource Sharing for Domain-Randomized Embodied Simulation}

\author{
{\rm Zixu Hao\textsuperscript{1}\quad
Weijun Wang\textsuperscript{1}\quad
Xiangyu Li\textsuperscript{1}\quad
Liang Mi\textsuperscript{1,2}}\\
{\rm Hao Wu\textsuperscript{2}\quad
Yunxin Liu\textsuperscript{1}\quad
Ting Cao\textsuperscript{1}}\\
{\small \textsuperscript{1}Institute for AI Industry Research (AIR), Tsinghua University}\\
{\small \textsuperscript{2}Nanjing University}
} % end author

\maketitle

%-------------------------------------------------------------------------------
\begin{abstract}
Simulation provides a low-cost sandbox where embodied agents can attempt tasks, observe outcomes, and learn from experience.
Running many diverse simulation environments concurrently can consume substantial CPU and GPU memory because independently constructed scenes duplicate resources that remain unchanged across instances.
We present \sys, a resource-sharing system for domain-randomized simulation, built around the representation $E=T+\Delta$: a shared template contains common resources, and a patch records the mutations needed to construct a target environment.
\sys separates randomization sampling from replay and selects reusable templates together with patches that preserve each target environment's intended configuration.
Its layered runtime shares compiled model structure and immutable assets while keeping mutable parameters and simulation state private to each instance.
We implement \sys on robosuite and MuJoCo and evaluate it on LIBERO-plus and RoboCasa365 workloads.
At 32 replicas, shared-template execution reduces CPU RSS by 88--92\% relative to process-per-environment execution.
Controlled experiments show how sharing shifts memory growth toward template diversity and unique asset content rather than instance count, within the supported sharing boundaries.
\end{abstract}

\section{Introduction}
\label{sec:introduction}

% P1 (agentic robotics)
Language models are expanding how robots can interpret instructions and carry out tasks in the physical world.
Recent work uses language models to generate and refine robot-control programs~\cite{code-as-policy,aspire}, control robots directly through action interfaces~\cite{robodawn,astra-robot-manipulation,astra-early-eval}, and coordinate vision-language-action (VLA) policies within agent harnesses~\cite{harnessvla,zetta}.
These systems open a path for embodied agents to improve through experience, making access to environments for repeated interaction an important part of their development.

% P2 (simulation as sandbox)
Simulation offers a low-cost \textbf{sandbox for embodied agents} to interact with their surroundings and learn from this interaction.
In Zetta, for example, repeated simulation rollouts supply failure traces for diagnosis, followed by further trials to validate proposed improvements~\cite{zetta}.
Physical trials require robot hardware, operating time, and supervision; simulated environments can be reset and run in parallel at scale with much lower costs.
Recent agentic real-to-sim-to-real systems reconstruct interactive simulation environments from real-world observations and transfer simulation-acquired skills or policies back to physical robots, making simulation a more practical environment for learning from experience~\cite{real2gym,arstag,robosimstudio}.
We see simulation as a sandbox for embodied agents, with the potential to become part of their post-training infrastructure.
The scale at which these sandboxes can run affects how much experience an agent can collect within a fixed time and hardware budget.
% \hzx{The original version uses "determines" rather than "affects", but that could be a little bit over-claim.}

% P3 (experience diversity)
Useful experience also requires variation in the conditions an agent encounters.
Domain randomization varies properties such as lighting, textures, and object poses to expose policies to different conditions and support transfer to the real world~\cite{domain-randomization}.
For example, repeated attempts at a pick-and-place task may use different object positions and table textures while retaining the same robot and room geometry.
Collecting this experience at scale requires many such environments to run concurrently.
Their combined CPU and GPU memory footprint limits how many environments fit on a machine, making efficient deployment of diverse simulation sandboxes a practical concern.

\begin{figure}[t]
\centering
\includegraphics[width=\linewidth]{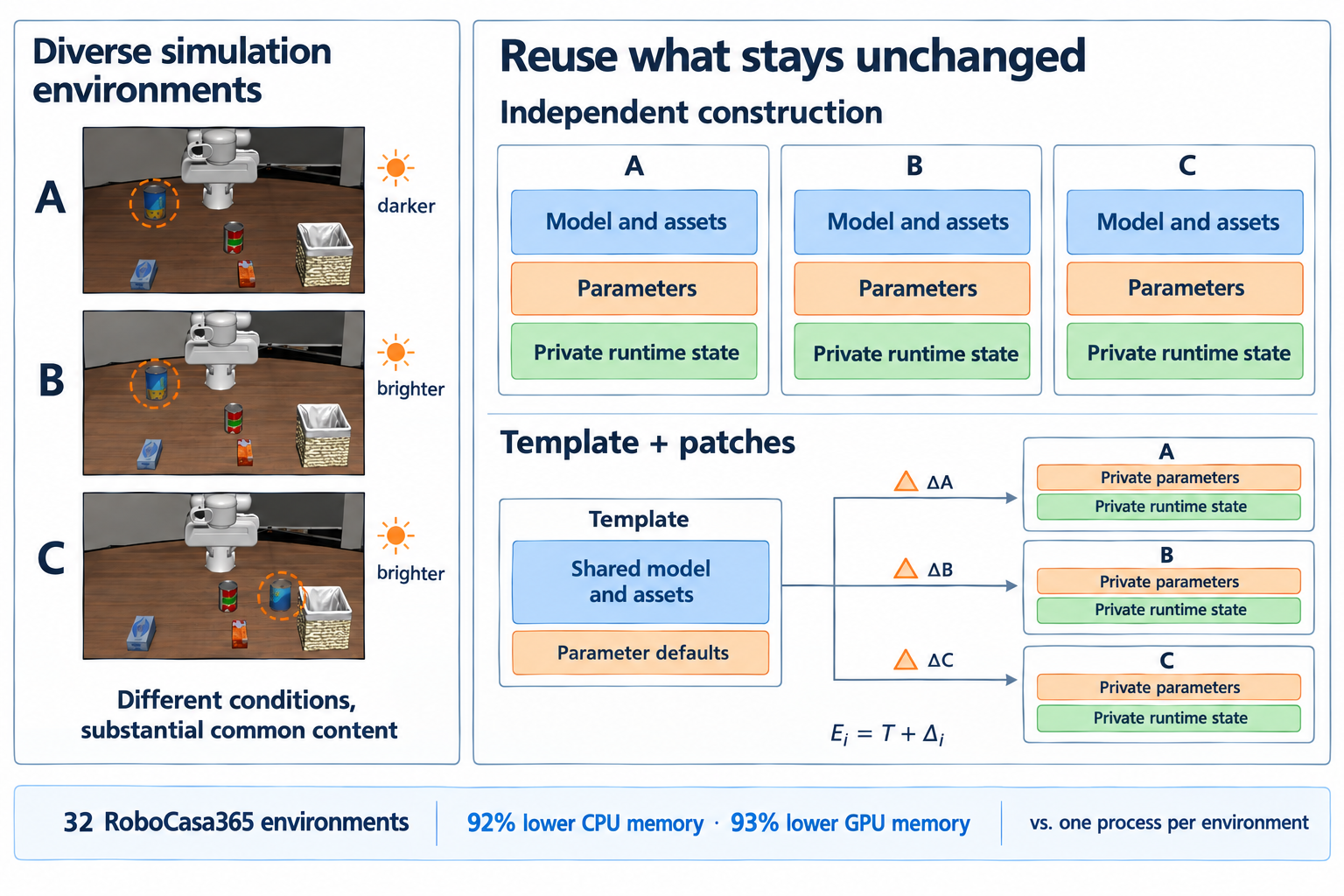}
\setlength{\abovecaptionskip}{3pt}
\caption{Resource sharing across diverse simulation environments.}
\label{fig:intro-sharing}
\end{figure}

% P4 (system issue)
Part of this memory footprint comes from duplicating resources that remain unchanged across diverse environments.
High-throughput simulators are optimized for running many homogeneous environments together.
\footnote{Recent Isaac Lab versions support heterogeneous cloning by routing asset variants to selected environments.
However, sharing remains bounded by USD prim/instance and backend-native physics boundaries; mutations that cross these boundaries may require uninstancing and prevent effective sharing.}
When environments differ, constructing separate models or scenes can copy common resources along with the properties that vary.
% For example, environments with different object configurations may still use the same robot geometry and textures, yet constructing an independent model for each can duplicate these resources.
ManiSkill's GPU path shares a PhysX system and simulation buffers while constructing heterogeneous sub-scenes independently~\cite{maniskill3}.
Many CPU-based MuJoCo stacks maintain a full model per worker, as in the subprocess environments used by RLinf for RoboCasa and LIBERO~\cite{rlinf,rlinf-osdi26}.
When independent models are materialized, model and asset resources grow with the number of environments, limiting how many simulation sandboxes can run concurrently on a machine.

% P5 (observation)
Many domain-randomized environments differ in only a small portion of their description.
A scene may retain the same robot, room geometry, and most assets while changing a light parameter, an object pose, or a texture reference.
These limited differences leave substantial common content that could be shared across environments.
Figure~\ref{fig:intro-sharing} illustrates this sharing opportunity.

% P6 (abstraction)
We represent this structure explicitly as $E = T + \Delta$.
A simulation environment $E$ consists of a shared template $T$ containing its invariant content and a patch $\Delta$ recording the mutations that produce one instance.
This representation allows environments to share unchanged resources while keeping the parts that vary private to each instance.

% P7 (challenge 1)
Turning this representation into efficient resource sharing raises three challenges.
First, simulator resources have no uniform structure or sharing granularity.
Compiled model structure, model parameters, runtime state, and rendering assets have different storage layouts, lifetimes, and ownership rules.
A model parameter may vary at the level of an individual field, while an immutable texture can be shared as a complete asset.
The system must map the mutations in a patch to sharing boundaries appropriate for each resource type.

% P8 (challenge 2)
Second, constructing an environment from a template and patch must preserve isolation and simulation semantics.
Each mutation in the patch must be applied to resources owned by the target instance, leaving the shared template and other instances unaffected.
The resulting environment must behave as if it had been constructed independently with the same specification.

% P9 (challenge 3)
Third, template reuse must preserve the intended randomization choices.
Simulation frameworks often sample these choices during construction or reset.
Reusing a template can bypass some of these steps or repeat others, producing a different environment if the choices remain implicit.
The system must record the choices and replay their effects so that applying a patch reproduces the target instance.

% P10 (overall design)
We present \sys, a simulation system that supports resource sharing for heterogeneous domain-randomized workloads.
It describes environment differences through a declarative mutation interface and uses a layered runtime to construct target instances from shared templates and patches.

% P11
To handle different resource granularities, \sys maps each supported mutation to the sharing boundary of the resource it affects.
The runtime shares compiled structure through templates, keeps mutable model parameters and runtime state private to each instance, and deduplicates immutable assets by content.

% P12
To preserve isolation during instance construction, \sys keeps templates read-only and directs supported patch mutations to instance-owned resources.
During template selection and planning, it compares structural and asset profiles to identify compatible templates and derive patches within the supported mutation vocabulary.

% P13
To preserve randomization choices across template reuse, \sys separates sampling from application and records the sampled choices as declarative mutations.
Instance construction replays these mutations without resampling their values.
This makes the choices explicit and replayable, allowing template reuse to reproduce the intended target instance.

% results
Our evaluation shows that, for compatible task variants, common model and asset resources need not be replicated per instance: model storage is shared within template groups, and texture storage is reused across groups.

We summarize our contributions as follows:
\begin{itemize}[nosep, leftmargin=*]
\item We introduce a template-and-patch representation for domain-randomized simulation environments, with explicit, replayable mutations describing the differences between a shared template and each target instance.
\item We design and implement \sys on the robosuite/MuJoCo stack. Its layered runtime shares compiled model structure and immutable assets while isolating per-instance model parameters and simulation state.
\item We evaluate \sys on robosuite-based LIBERO-plus and RoboCasa365. With 32 RoboCasa365 environments, it reduces CPU RSS and GPU memory by 92\% and 93\%, respectively, compared with process-per-environment execution.
\end{itemize}

\section{Background}
\label{sec:background}

\subsection{Simulation Software Stack}

Embodied simulation stacks combine task definitions, an environment framework, and physics and rendering backends.
Table~\ref{tab:sim-software-stacks} lists representative stacks.
LIBERO and RoboCasa use robosuite with MuJoCo.

MuJoCo provides physics simulation and rendering interfaces.
robosuite adds environment construction and controllers for multiple robot embodiments.
It also provides interfaces for advancing simulations and obtaining observations and rewards.
LIBERO defines manipulation tasks on this stack, and LIBERO-plus~\cite{libero-plus} extends them with generated task variants.
RoboCasa constructs household scenes and tasks procedurally; RoboCasa365 provides a benchmark of 365 everyday tasks.\footnote{RoboCasa (v0.2) and RoboCasa365 (v1.0 onward) are releases of the same project. We distinguish them because their task sets and asset libraries differ substantially.}

\begin{table}[t]
\centering
\small
\setlength{\tabcolsep}{4pt}
\caption{Representative embodied simulation software stacks.}
\label{tab:sim-software-stacks}
\begin{tabular}{lll}
\toprule
Sim. Env. & Framework & Physics / Rendering \\
\midrule
LIBERO~\cite{libero} & \multirow{2}{*}{robosuite~\cite{robosuite}} & \multirow{2}{*}{MuJoCo~\cite{mujoco} / OpenGL} \\
RoboCasa~\cite{robocasa} & &\\
\midrule
ManiSkill~\cite{maniskill2,maniskill3} & SAPIEN~\cite{sapien} & PhysX~\cite{physx} / Vulkan \\
\midrule
Isaac Lab~\cite{isaaclab} & Isaac Sim~\cite{isaacsim} & \multirow{2}{*}{PhysX / RTX} \\
BEHAVIOR-1K~\cite{behavior} & OmniGibson & \\
\bottomrule
\end{tabular}
\end{table}

\subsection{MuJoCo Models and Simulation State}

MuJoCo separates a compiled scene description from the state of a simulation instance.
It compiles an XML scene description into an \texttt{mjModel}, which contains the scene's bodies, geometries, joints, actuators, assets, and simulation parameters.
An \texttt{mjData} object stores the evolving state of one instance, including positions, velocities, contacts, sensor readings, and solver state.

\begin{figure}[t]
\centering
\begin{tikzpicture}[
  font=\small,
  x=1in,
  y=1in,
  model/.style={
    draw,
    rounded corners=1pt,
    minimum height=0.38in,
    minimum width=1.35in,
    align=center
  },
  data/.style={
    draw,
    rounded corners=1pt,
    minimum height=0.35in,
    minimum width=0.92in,
    align=center
  },
  flow/.style={->, >=stealth, thick},
  note/.style={
    draw,
    dashed,
    rounded corners=1pt,
    inner sep=4pt,
    align=center,
    font=\scriptsize
  }
]
  \node[model] (xml) at (2.05, 0) {XML\\scene description};
  \node[model] (model) at (2.05, -0.58) {\texttt{mjModel}\\compiled scene};
  \draw[flow] (xml) -- node[right, font=\scriptsize] {compile} (model);

  \node[data] (data1) at (0.95, -1.1) {\texttt{mjData\textsubscript{1}}\\runtime state};
  \node[data] (data2) at (2.05, -1.1) {\texttt{mjData\textsubscript{2}}\\runtime state};
  \node[data] (data3) at (3.15, -1.1) {\texttt{mjData\textsubscript{3}}\\runtime state};

  \draw[flow] (model.south) -- (data1.north);
  \draw[flow] (model.south) -- (data2.north);
  \draw[flow] (model.south) -- (data3.north);

  % \node[note] at (2.05, -1.55)
  %   {Model-level differences require additional\\\texttt{mjModel} instances};
\end{tikzpicture}
\caption{MuJoCo separates the compiled model (\texttt{mjModel}) from
per-instance runtime state (\texttt{mjData}).}
\label{fig:mujoco-model-data}
\end{figure}
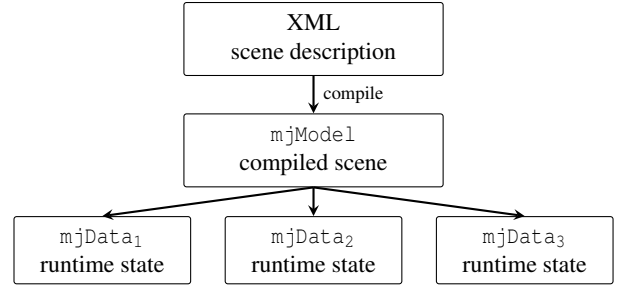

During physics simulation (\texttt{mj\_step}), MuJoCo treats \texttt{mjModel} as read-only and updates the instance's \texttt{mjData}.
An unchanged model can therefore be shared by multiple instances, each advancing independently with its own \texttt{mjData} object (Figure~\ref{fig:mujoco-model-data}).

The compiled model uses a structure-of-arrays representation with fixed-size storage allocated during compilation.
Structural changes, such as adding bodies or geometries, change the array layout and require recompilation.
Some parameter values, such as lighting, can instead be edited in an existing model without changing its layout.
If concurrent instances require different values for these parameters, they cannot use the same model unchanged: writing a shared model field affects every instance using it.
Running these variants with separate models also duplicates the unchanged parts of those models.

\subsection{Domain Randomization Interfaces}

\begin{figure}[t]
\centering
\includegraphics[width=\linewidth]{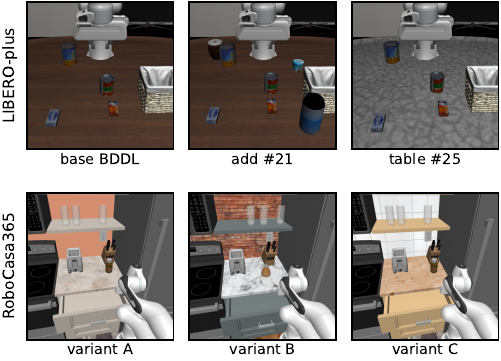}
\caption{Domain randomization examples from LIBERO-plus and RoboCasa365.}
\label{fig:randomization}
\end{figure}

Applications express randomization choices through different interfaces: task-description files, environment construction options, or direct model edits.
These interfaces also differ in when the choices take effect.

LIBERO describes tasks in BDDL, a declarative task-description format.
LIBERO-plus generates variants of these descriptions offline, for example by adding objects or changing table appearances, and stores each variant as a separate BDDL file.
These choices enter the environment before the scene is compiled.

RoboCasa makes choices during environment construction, selecting layouts, styles, and objects and sampling their placements.
The texture variants in Figure~\ref{fig:randomization} use different texture sets within the same layout and style.

robosuite also exposes \emph{modders} that edit fields of an existing model.
They provide interfaces for changing lighting, cameras, materials and textures, and dynamics parameters such as friction, mass, and damping.
These edits change model data directly rather than rebuilding the scene from its task description.

\section{Characterizing Domain-Randomized Workloads}
\label{sec:characterization}

In this section, we characterize the domain-randomized workloads and derive the requirements they place on the system design.

\subsection{Variation Is Sparse and Structured}

\begin{figure*}[!t]
\centering
\includegraphics[width=0.85\textwidth]{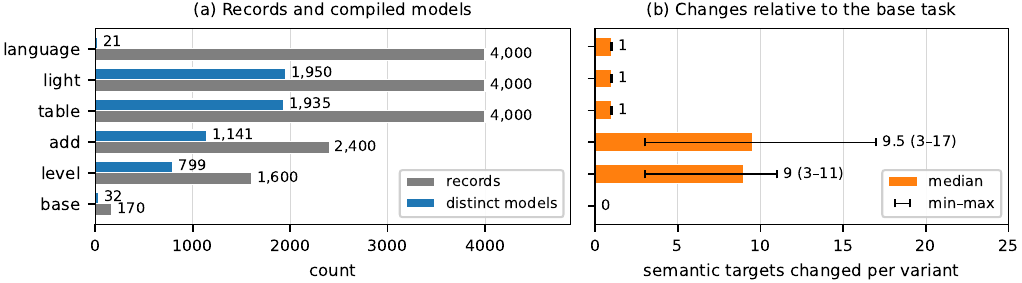}

\includegraphics[width=0.97\textwidth]{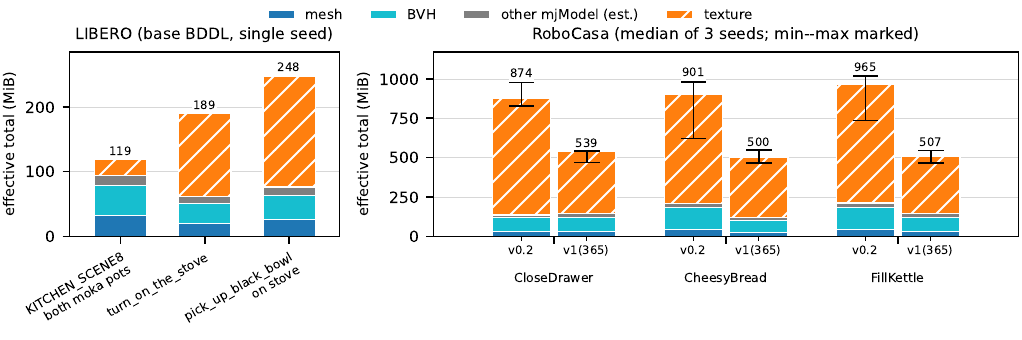}
\caption{Top: compiled models (a) and per-variant changes (b) in LIBERO-plus.
Bottom: model and asset memory in LIBERO and RoboCasa.
RoboCasa memory bars show medians of three seeds with min--max markers.}
\label{fig:sec03-characterization}
\end{figure*}

LIBERO-plus provides an enumerable set of task variants.
Across 16,170 task records, we identify 5,855 distinct compiled models (Figure~\ref{fig:sec03-characterization}a).
We group records by scene and construction metadata, then verify within-group model equivalence by compiling every member and comparing its model profile.
For example, 4,000 language variants map to 21 models, all of which also occur in other families.
Changing the instruction does not require a different model.

Variation is also concentrated in explicit targets.
Across the 16,000 non-base variants, 75\% change just one semantic target in the task description: lighting, language, or background texture (Figure~\ref{fig:sec03-characterization}b).
Object-placement perturbations (\texttt{level}) and distractor additions (\texttt{add}) affect a broader set of targets, but no variant changes more than 17.
The same qualitative pattern appears in other interfaces.
robosuite modders organize changes into lighting, camera, texture, and dynamics fields, while RoboCasa exposes choices for layouts, styles, objects, and placements.
These interfaces generate diverse environments through explicit, structured choices.

\subsection{Memory Footprints Are Layered and Skewed}

Memory costs are unevenly distributed across an environment's resources.
A compiled model contains large assets such as textures and meshes alongside smaller structural arrays and simulation parameters.
Environments can differ in a few parameters while retaining the same large assets.
Copying the whole model to isolate those parameters also duplicates the unchanged assets.

The bottom panel of Figure~\ref{fig:sec03-characterization} decomposes this logical footprint for three LIBERO tasks and three RoboCasa tasks.
Texture pixels alone account for 20--69\% of the LIBERO models.
For \texttt{CloseDrawer}, textures account for 81\% of the footprint in RoboCasa v0.2 and 70\% in RoboCasa365.
The corresponding model and asset footprints are 874~MiB and 539~MiB.

Smaller components still need isolation when their values differ across environments.
Parameters such as light, camera, body, and site fields reside in the compiled model, so writing them in a shared model affects every instance using it.
Each instance has its own \texttt{mjData}, whose main buffer occupies less than 1\% of the model and asset footprint, reaching at most 0.77~MiB for the measured LIBERO tasks and 2.07~MiB for RoboCasa365.
Sharing must preserve reusable model and asset data while keeping environment-specific parameters and runtime state isolated.

\subsection{Coupling Randomization with Construction Undermines Sharing and Replayability}

The interfaces described in Section~\ref{sec:background} apply variation during environment construction or through edits to an existing model.
Construction typically produces a complete model for each variant, duplicating unchanged resources alongside the differences.
Editing model fields avoids rebuilding the model, but those writes affect every instance that shares it.
Supporting variation efficiently therefore requires identifying which resources must be constructed or privately owned for each change.

RoboCasa illustrates the cost of applying these decisions at the environment level.
Its default hard-reset path reconstructs the scene and reinitializes the simulator even when changes are confined to model parameters or simulation state.
Reusing an existing model avoids this work but can skip randomization choices embedded in construction.
In our RoboCasa365 integration, we found that the reconstruction path mixed restored episode metadata with freshly sampled task-specific choices.
Such partial replay can silently produce an environment that no longer matches the recorded episode, undermining comparisons intended to use the same environment.
Sharing therefore requires both selective materialization and faithful replay of the choices that define each target environment.

\subsection{Workload Properties and Requirements}

The workloads examined here exhibit sparse, structured variation across resource layers with different memory costs and sharing opportunities.
These observations motivate representing each environment as a reusable template and an explicit patch, $E = T + \Delta$.
Supporting this decomposition imposes four requirements.

\textbf{Explicit variation.}
The representation must identify the targets and values of each change across application interfaces.
\textbf{Resource-specific sharing.}
Materialization must reuse unchanged resources at the granularity supported by their storage and access patterns.
\textbf{Isolation and semantic equivalence.}
Constructing one instance must leave shared resources and other instances unchanged, and produce the same model and behavior as independent construction of the target environment.
\textbf{Faithful replay.}
Reconstruction must preserve sampled choices without introducing additional randomization.

%-------------------------------------------------------------------------------
\section{Design Overview}
\label{sec:design-overview}
%-------------------------------------------------------------------------------

\begin{figure*}[t]
\centering
\includegraphics[width=\textwidth]{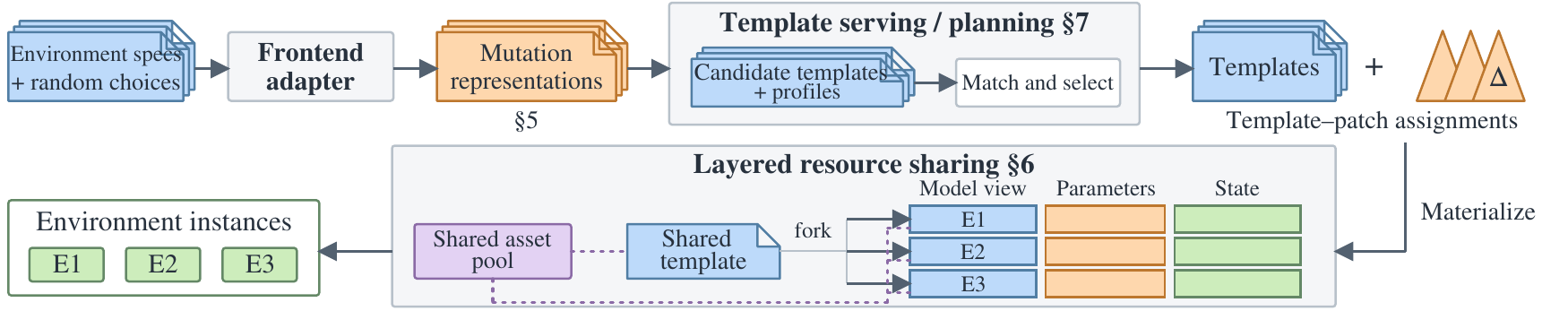}
\caption{Overview of \sys. The pipeline transforms environment specifications into mutation representations,
  serves them with shared templates and patches, and materializes them through layered resource-sharing mechanisms.}
\label{fig:overview}
\end{figure*}

Figure~\ref{fig:overview} shows the execution pipeline of \sys.
Given environment specifications and randomization choices, \sys constructs a common mutation representation (§\ref{sec:mutation-ir}).
A serving policy selects reusable templates and represents each target environment as a patch over one template (§\ref{sec:template-serving}).
The resulting template--patch pairs are materialized through layered sharing mechanisms, which share compiled structure and immutable assets while isolating per-environment parameters and runtime state (§\ref{sec:sharing-mechanism}).

We first describe the mutation representation and the underlying sharing mechanisms, then explain how template serving connects the two.
Section~\ref{sec:mutation-ir} makes variation explicit and replayable, exposing the differences that determine sharing opportunities.
Section~\ref{sec:sharing-mechanism} realizes these opportunities through resource-specific mechanisms that preserve instance isolation and simulation semantics.
Section~\ref{sec:template-serving} bridges the representation and these mechanisms by discovering reusable templates and deriving the patches that construct each target environment from them.

\section{Mutation Representation and Materialization Contract}
\label{sec:mutation-ir}

\subsection{Mutation Specifications}

We represent one environment change as an immutable \texttt{MutationSpec}.
Each specification names a logical target, an operation, and a value:

\begin{lstlisting}
target = MutationTarget("object", ("moka_pot_1", "pose"))
spec = MutationSpec(target, "set", pose)
\end{lstlisting}

The target is a stable namespace and path, such as
\texttt{object/moka\_pot\_1/pose} or
\texttt{scene/light/0/diffuse}.
It does not expose an engine-specific storage location.
The generic representation admits \texttt{set}, \texttt{add}, and
\texttt{remove}.
The backend-specific vocabulary restricts these operations
to targets that the resolver can materialize.
For the current MuJoCo backend, compile-free serving primarily uses isolated
\texttt{set} operations.
The specification records what changes, but does not state whether the change
requires a new compiled template, instance-owned model storage, or per-environment
state.

Benchmark and framework frontends generate this common representation for
storage, template planning, and instance construction.
Canonical serialization preserves each specification's target, operation, and
value across these modules.
Mutation specifications thus serve as an intermediate representation (IR)
behind the existing application APIs.

\subsection{Sampling and Replay}

Sampling produces mutation data, while application consumes that data.
The sampler receives a local random-number generator and returns an ordered
sequence of specifications:

\begin{figure}[t]
\centering
\includegraphics[width=\columnwidth]{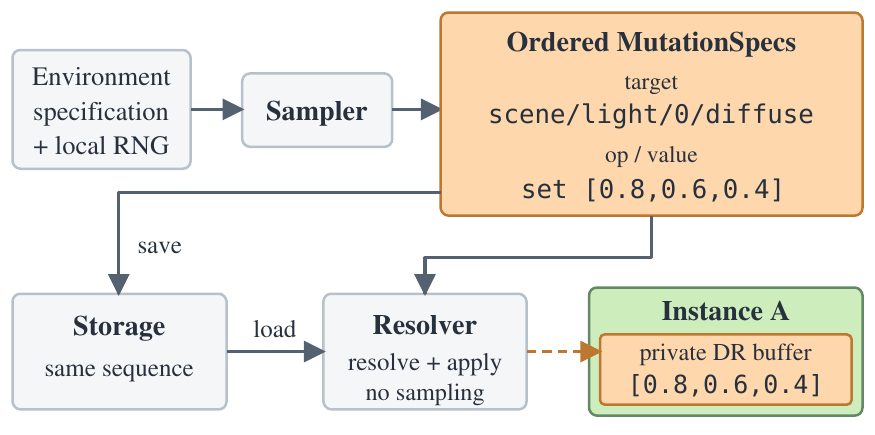}
\vspace{-4pt}
\caption{Sampling produces an explicit mutation sequence that can be saved
and replayed without further randomness.}
\label{fig:mutation-lifecycle}
\end{figure}

\begin{lstlisting}
class LightSampler:
    def sample(self, rng):
        value = tuple(rng.uniform(0.0, 1.0, 3))
        target = MutationTarget("scene", ("light", "0", "diffuse"))
        return (MutationSpec(target, "set", value),)
\end{lstlisting}

The sampler may use the generator for collision checks or retries, but those
choices remain inside \texttt{sample}.
Applying a sampled sequence uses a resolver that maps each logical target to a
backend-specific writer:

\begin{lstlisting}
mutations = sampler.sample(np.random.default_rng(seed))
apply_mutations(env, mutations, resolver=resolver)
\end{lstlisting}

\texttt{apply\_mutations} consumes no random numbers and preserves the order
of the sequence.
The same sequence can therefore be stored, applied to another compatible
environment, or encoded as part of a patch and replayed later.
The recorded sequence can be attached to a selected template and applied during
materialization without resampling.

\subsection{Template and Patch Semantics}

Once a template is fixed, the mutations that distinguish a target environment
from that template form its patch.
We represent the result for environment $E_i$ as a template identifier and an
ordered patch:
\[
    E_i = T_{g(i)} + \Delta_i,
\]
where $g(i)$ selects the template and $\Delta_i$ is a sequence of
\texttt{MutationSpec} objects.
The sequence can be persisted with \texttt{encode\_patch} as a versioned
record containing the template key and its mutations.
The patch is valid only when every mutation has a defined interpretation
relative to the selected template.
Section~\ref{sec:template-serving}
describes how the system makes that selection.

The central contract is that materializing $\Delta_i$ over $T_{g(i)}$ produces
the same model and execution behavior as constructing $E_i$ independently.
Section~\ref{sec:sharing-mechanism} describes how the runtime preserves this
contract while sharing structure and assets.

\subsection{Logical Targets and Resource Mapping}

Mutation targets identify logical environment fields, while the resolver maps
them to the resource and execution stage that owns the corresponding write.

\begin{table}[t]
\centering
\small
\setlength{\tabcolsep}{3pt}
\caption{Logical mutation targets and their materialization destinations.}
\label{tab:mutation-targets}
\begin{tabular}{p{0.21\columnwidth}p{0.22\columnwidth}p{0.22\columnwidth}p{0.16\columnwidth}}
\toprule
Target & Example & Destination & Owner \\
\midrule
Structure & object addition & model arrays & template \\
Model field & light diffuse & DR buffer & instance \\
Runtime state & object pose & \texttt{mjData} & instance \\
Metadata & task language & Python env & instance \\
\bottomrule
\end{tabular}
\end{table}

In the current backend, changes to structure or asset membership require building a new template,
which remains read-only once shared.
Model-field changes write to instance-owned DR buffers.
Poses and other dynamic values are applied to the instance's \texttt{mjData},
and task metadata is handled by its Python environment.

This mapping is deliberately outside \texttt{MutationSpec}.
The same logical target can therefore be interpreted by a backend-specific
resolver without changing the serialized representation, while the resolver
can reject an operation that its target does not support.
The interface records the mutation and its order.
Template selection and the execution path determine where that mutation is
materialized.

\section{Layered Model Resource Sharing}
\label{sec:sharing-mechanism}

Section~\ref{sec:mutation-ir} defines a patch as an ordered sequence of logical
writes over a selected template.
This section describes how \sys materializes those writes without copying
every resource for every environment.
The runtime uses separate mechanisms for compiled model structure, mutable
model parameters, and immutable assets.  The mechanisms share a common rule:
the template remains read-only, and every write is directed to the owner of
the target environment.

\begin{figure*}[t]
\centering
\includegraphics[width=\textwidth]{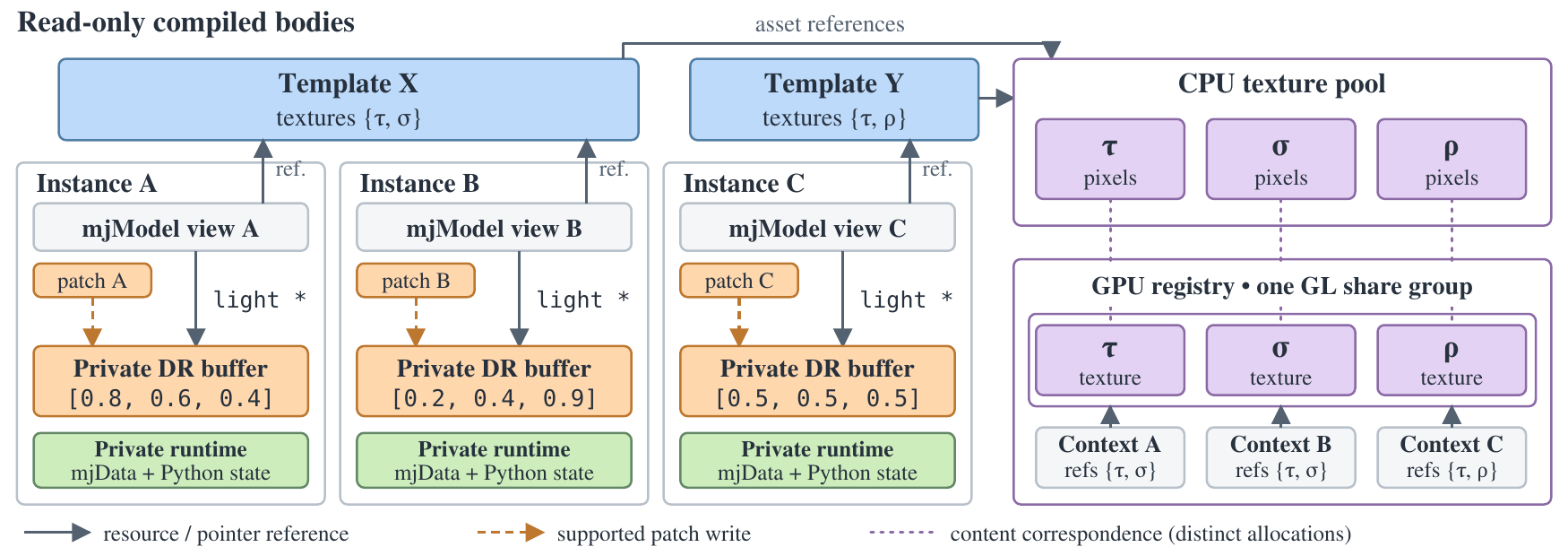}
\vspace{-4pt}
\caption{Compiled structure is shared within template groups, while immutable
textures are deduplicated across groups, separately on CPU and GPU.}
\label{fig:layered-sharing}
\end{figure*}

\subsection{Structure Layer: Model Forking}

MuJoCo allows instances to share an unchanged \texttt{mjModel} while
maintaining separate \texttt{mjData} objects.
Instances that require different model-field values need separate writable
storage.
\sys adds a model-level fork operation that creates an environment-specific
model view while reference-counting the large read-only body of the compiled
model.
The fork owns its writable model metadata and its \texttt{mjData}.
Unchanged compiled structure remains shared with the template.

The fork operation must reproduce the complete simulation environment, not
only the MuJoCo model.
Robosuite also holds Python-side construction state, including choices that
determine the generated MJCF and the initial simulation state.
The template provides a baseline for homogeneous copies: a compiled model,
a deterministic initial state, and metadata needed to replay construction
decisions.
The capture path disables reset-time sampling but retains the structural
choices that determine the compiled model.
A fork rebuilds private Python objects from the captured construction
metadata and creates a shallow model copy with its own \texttt{mjData}.
The captured state provides a common baseline for instance-specific model
and state changes, allowing compatible variants to reuse the compiled model
while preserving their intended differences.

Process-level copy-on-write shares memory at page granularity, without
distinguishing model structure, mutable parameters, and renderer resources.
It does not specify how Python environments and graphics contexts should be
reconstructed for each instance.
The model-level operation instead makes the shared body and the per-environment
owners explicit.
The implementation details of its reference counting and
backend integration appear in Section~\ref{sec:implementation}.

\subsection{Parameter Layer: Fork-Local DR Buffers}

Some mutations change model values after compilation, such as light, camera,
body, and site parameters.
These fields cannot be written into the shared model body because a write for
one environment would change every environment using that template.
For these fields, each fork has a small private DR buffer, and the relevant
model pointers resolve to that buffer after the fork.
The engine can then use its normal field accesses while the mutation resolver
writes only to the fork-local copy.

Each fork receives a deep copy of the DR buffer, giving it independent
storage for mutable model parameters.
The buffer is small compared with the shared model and asset data, making a
deep copy cheaper and easier to reason about than a general copy-on-write
scheme.

The DR boundary depends on both storage cost and update propagation.
Small parameter arrays are inexpensive to copy, but their updates must also
preserve the engine's derived state.
The backend admits a field only when its writes and any required derived
updates stay within instance-owned storage.
Fields whose updates would affect shared storage are excluded from the DR
vocabulary and require a different template or materialization path.
The backend explicitly defines this propagation contract for each supported
field.
Page-level copy-on-write can isolate modified bytes, but does not trigger the
application-specific updates needed to keep related fields and derived state
consistent.

\subsection{Asset Layer: Content-Addressed Deduplication}

Meshes and textures are immutable after loading and can be shared across
models even when their surrounding environments differ.
\sys assigns each asset a content identity and keeps shared references in a
process-wide CPU pool.
Copying a model increments references to existing asset data instead of
copying its bytes.
The renderer applies the same idea to GPU texture objects across the workers'
shared GL context group, using content together with texture metadata as the
identity.
The CPU and GPU pools are separate because they own different resource types,
but both implement the same rule: equal immutable content has one physical
copy and each model or context holds an explicit reference to it.

\section{Template Selection and Serving}
\label{sec:template-serving}

Section~\ref{sec:mutation-ir} defines a served environment as a template and
an ordered patch, and Section~\ref{sec:sharing-mechanism} describes how the
pair is materialized.
This section describes the policy layer that connects the two: given a set of
environment specifications, it decides which templates to make available and
which patch serves each specification.
The policy may resolve one specification against templates already in a cache,
or plan a complete batch before any new template is built.
In either case, its output assigns every environment
to one template and one patch.

\begin{figure*}[t]
\centering
\includegraphics[width=\textwidth]{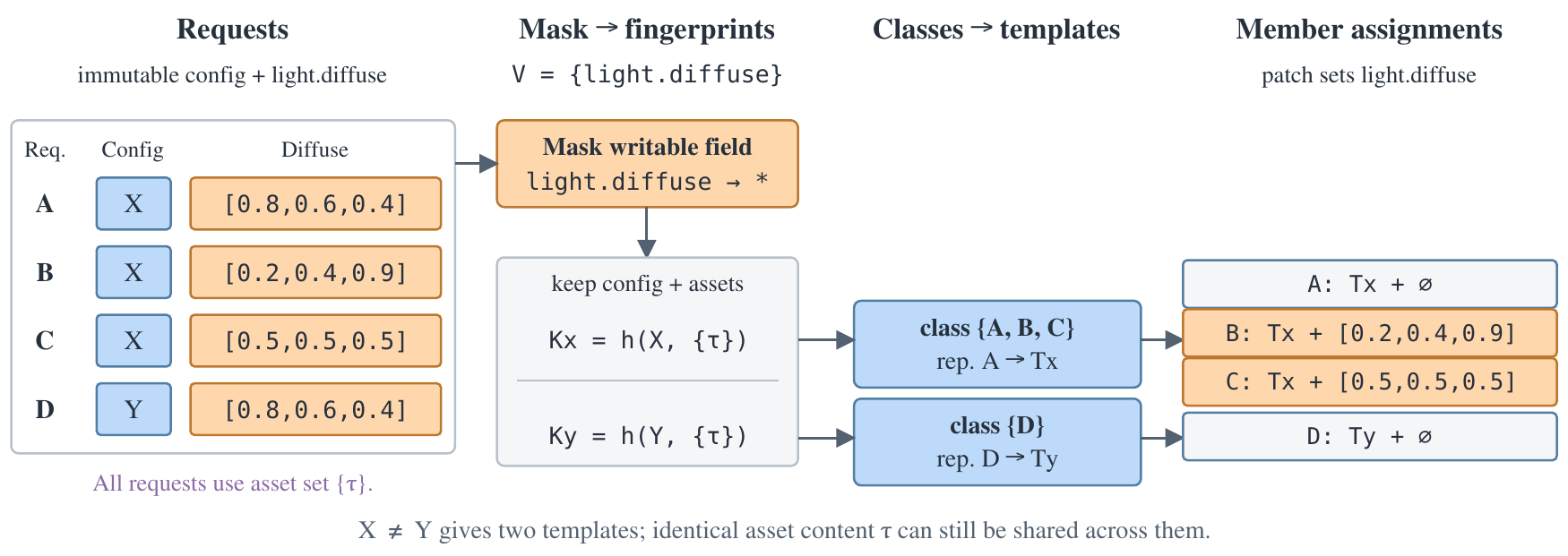}
\caption{Template serving groups profiles by their unchanged structure and
asset identities, selecting one representative per class and deriving member patches.}
\label{fig:equivalence-serving}
\end{figure*}

\subsection{Incremental Selection and Batch Planning}

\sys exposes two planning interfaces with the same output contract.
Both use profiles summarizing model structure, parameter values, and immutable
asset identities to check compatibility and derive patches.
The frontend supplies these profiles without exposing its construction path
to the selection policy.

Incremental selection handles an arriving specification using the templates
already registered in a cache.
It obtains the specification's profile, asks a selection policy for a
compatible source, and returns a \texttt{TemplateAssignment} containing the
source id and the derived patch.
On a miss, the cache registers the specification as a new source rather than
forcing it into an incompatible template.
This path is the generic fallback because it does not require the workload to
be known in advance.

Batch planning receives the whole collection of specifications and returns a
\texttt{BatchTemplatePlan}:

\begin{lstlisting}
plan = cache.plan(
    members, profile_of=profile_of,
    vocabulary=vocabulary, policy=policy)
\end{lstlisting}

The plan contains one build record for each template that must be materialized
and one assignment for each workload member.
The policy chooses how to group profiles, which member anchors each group, and
which patch derives every other member from that anchor.
The cache applies the plan only after planning completes, so template
construction can be performed once per selected source and in parallel when
the caller provides a parallel materialization path.

Batch planning is optional.
The default profile-based policy composes incremental selection with grouping
and one anchor per remaining group.
A workload-specific policy can replace
that strategy while preserving the same assignment and build-plan interface.
The current LIBERO-plus offline index is a static instance of this batch path:
it names representative sources and their patches before runtime construction,
while unresolved requests can still fall back to the incremental cache.

\subsection{The Operational Serving Preorder}

Let $V$ be the backend's compile-free materialization vocabulary, rather than
all operations expressible by the generic mutation representation.
We say that $B$ serves $A$ when a
valid ordered patch from $V$ materializes $A$ from $B$:
\[
  A \preceq_V B
  \quad\Longleftrightarrow\quad
  \exists\Delta\in V^*:\;
  \operatorname{apply}(B,\Delta)\equiv A.
\]
The equivalence here is the materialization contract of
Section~\ref{sec:mutation-ir}, including the environment state and behavior.
The patch must preserve the write-isolation requirements of
Section~\ref{sec:sharing-mechanism}.  The direction is from the target $A$ to
the template $B$ that can serve it.

An empty patch gives $A\preceq_V A$.  If valid patches can be composed without
recompilation, concatenating patches from $C$ to $B$ and from $B$ to $A$ gives
$A\preceq_V C$.
Under this closure condition, $\preceq_V$ is a preorder.
Its mutual-serving kernel,
$A\sim_V B$ iff $A\preceq_V B$ and $B\preceq_V A$, is an equivalence
relation.  A one-way serving relation can cover several targets with one
template, whereas the kernel partitions environments into classes whose
members can serve one another.

The preorder defines a larger search space for batch planning.
The current policy uses its mutual-serving kernel under the restricted MuJoCo
vocabulary: it partitions profiles into
equivalence classes and selects one representative for each class.
A more general policy could select templates whose serving regions cover the
whole workload.  For example, if the mutation vocabulary can select asset
bindings, a policy could construct a union template for several classes and
serve each member with a class-specific patch.
This policy would require executable support for the additional mutations and
is outside the current implementation.

To explore such union templates, a policy can use a structural--asset cover
test as a preliminary filter before deriving executable patches.
The test checks for compatible non-asset topology, sufficient reachable asset
counts, and an asset set covering the target's requirements.
Passing this test identifies a candidate whose resources could cover the
target, subject to backend support for selecting asset bindings and applying
the required writes.

\subsection{Equivalence-Based Serving}

The current implementation instantiates batch planning with the mutual-serving
kernel of the operational preorder.
Two profiles belong to the same class when they have identical immutable asset
identities and agree outside the structure fields covered by the mutation
vocabulary.
The class fingerprint excludes patchable fields from the structural comparison
and includes the sorted asset identities.
Within each class, the policy selects one representative and derives the
field-level patches for the other members:

\begin{lstlisting}
vocabulary = (PatchField(1, ("scene", "light"), "diffuse"),)
patch = derive_patch(template_profile, member_profile, vocabulary)
\end{lstlisting}

This policy turns planning into deterministic grouping followed by
representative selection.  The grouping is independent of input order, while
the representative can be chosen by a policy such as minimum total patch
writes.  The resulting assignments are recorded explicitly so that template
construction and environment creation consume the same plan.

\subsection{Offline Index Construction and Validation}

When a workload or a useful subset of it can be enumerated in advance, the
batch interface can build an offline index of template assignments.
The index records each covered specification's selected template and patch.
It can be reused across runs through the same assignment interface as the incremental cache.
Planning groups specifications using their profiles and selects the templates
to construct.
Requests outside the indexed subset use incremental selection, so the index
can be partial or omitted entirely.

The indexed path is summarized below.

\begin{lstlisting}
for member in workload:
    profiles[member] = profile_of(member)

groups = group_by_key(profiles, vocabulary)
plan = select_representatives(groups, policy)

for source in plan.templates:
    template = materialize(source.representative)
    check_profile(template, profiles[source.representative])
    for member in source.members:
        patch = derive_patch(
            profiles[source.representative], profiles[member], vocabulary)
        check_profile_patch(template, patch, profiles[member])
        index[member] = (source.id, patch)
\end{lstlisting}

The first check ensures that the profile used for planning describes the model
that is actually materialized.
The second checks that the template and patch reproduce the member's profile
within the supported mutation vocabulary.
The full materialization contract also requires agreement in derived state
and execution behavior with an independently constructed target environment.
Section~\ref{sec:evaluation} reports the coverage and results of profile and
execution-level validation.

\section{Implementation}
\label{sec:implementation}

We implement \sys on the MuJoCo/robosuite stack with approximately 18K LoC.

\subsection{MuJoCo Engine Extensions}

We add a shallow-copy fork primitive and use reference counting to manage the shareable body of an \texttt{mjModel}.
Each model can own a private DR buffer, which is deep-copied when the model is copied.
Enabling a DR group redirects the corresponding model pointers into this buffer, allowing existing engine field accesses to use instance-owned storage.
We extend the engine's macro-based field declarations to define DR groups and derive buffer sizing, field copying and pointer redirection.
Fields admitted to these groups must satisfy the propagation contract described in Section~\ref{sec:sharing-mechanism}.

The current asset-pooling implementation deduplicates textures on both CPU and GPU.
The CPU texture pool uses content hashes and reference counting to manage shared pixel data.
A \texttt{mjModel} no longer exposes \texttt{tex\_data} as one contiguous pixel array, and texture-size statistics report logical bytes referenced by the model.
A renderer-side registry shares GPU texture objects across GL contexts using content and texture metadata as the key.

\subsection{Threaded robosuite Runtime}

We run environments as threads within one process so they can access shared model and asset pools.
The Python controller path in robosuite limits concurrent execution through the GIL.
We move the physics and controller computations on the simulation hot path into C++, releasing the GIL during their execution.
Correctness checks compare control outputs and simulation states against the Python path from matching initial states and actions.

Environments are assigned to fixed worker threads, each owning a GL context.
The contexts of all workers belong to one EGL share group, enabling texture sharing in GPU memory.

\section{Evaluation}
\label{sec:evaluation}

We evaluate how sharing affects resource use and throughput, how resource use
changes with workload diversity, and how effectively templates serve related
environments.  We check semantic equivalence alongside the serving results.

\subsection{Evaluation Setup}

\paragraph{Workloads.} We use three workload families on the
robosuite/MuJoCo stack: the base robosuite benchmark suite, LIBERO-plus,
which provides an enumerable corpus of task variants, and RoboCasa365,
which supplies asset-rich household scenes.  Unless stated otherwise, the
workloads are robosuite's PickPlaceMilk, the libero\_10 living-room basket
task of LIBERO-plus, and RoboCasa365's CloseDrawer.

\paragraph{Software.} \sys is implemented on forked robosuite 1.5.2 and
MuJoCo 3.8.1 code bases, with the LIBERO-plus (v0.1.0) and RoboCasa365
(v1.0.1) benchmark packages on top.

\paragraph{Testbed.} Experiments run on a server with two Intel Xeon Silver
4314 CPUs (32 physical cores total) and NVIDIA A40 GPUs with 48\,GB of memory each.
Each GPU benchmark run uses one selected A40 for EGL offscreen rendering.

\paragraph{Metrics.} We report throughput in environment steps per second,
CPU RSS, and process-attributed GPU memory.
CPU RSS sums the resident set sizes of the benchmark process and its descendants.
GPU memory is attributed to those processes on the selected device.
Scaling experiments average memory samples taken during steady-state execution after warmup.
Scaling results report medians over three seeds, with throughput measured
after warmup.

\subsection{Resource Use and Throughput under Replica Scaling}
\label{sec:eval-scaling}

We first measure stepping with camera and object observations disabled in
\texttt{env.step}, while keeping offscreen rendering initialized and available
on demand.
This configuration excludes Python-side camera and object observation
processing, whose GIL contention limits threaded scaling.
We then enable camera observations to measure their additional cost.

\begin{figure*}[t]
\centering
\includegraphics[width=0.98\textwidth]{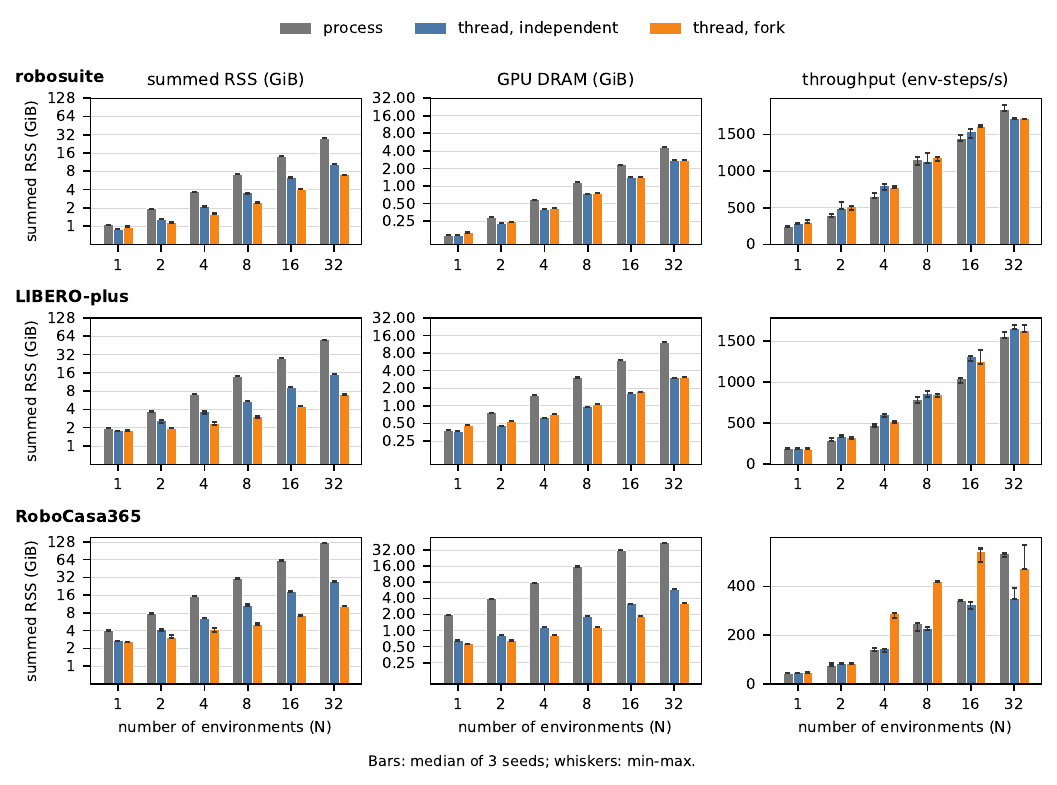}
\caption{Resource use and throughput across replica counts.
Bars show medians over three seeds with min--max markers.}
\label{fig:sec09-c2-scaling}
\end{figure*}

\paragraph{Replica scaling.} For each workload, we compare
process-per-environment execution, threaded execution with independent models, and
threaded execution with shared templates as the replica count increases.
Each comparison uses the same task and observation configuration.
GPU texture sharing is enabled in all three configurations.
Figure~\ref{fig:sec09-c2-scaling} shows substantial memory savings with
throughput close to process-per-environment execution.
At $N{=}32$, shared-template threads reduce summed CPU RSS by 88\% on
LIBERO-plus and 92\% on RoboCasa365 relative to separate processes.
On RoboCasa365, summed RSS falls from 123.8 to 10.5\,GiB, compared with
27.6\,GiB for independent threads.
GPU memory falls from 44.0 to 3.2\,GiB, compared with 5.9\,GiB for
independent threads.
Throughput is 1{,}632 versus 1{,}550 steps/s on LIBERO-plus and 475 versus
533 steps/s on RoboCasa365 for shared-template threads and separate
processes, respectively.

\begin{table}[h]
\centering
\small
\setlength{\tabcolsep}{3.6pt}
\caption{Effect of enabling camera observations on RoboCasa365.
Values are medians over three seeds.}
\label{tab:s2-camera}
\begin{tabular}{llrrrrrr}
\toprule
 & & \multicolumn{2}{c}{Steps/s} &
     \multicolumn{2}{c}{GPU (GiB)} &
     \multicolumn{2}{c}{RSS (GiB)} \\
\cmidrule(lr){3-4}\cmidrule(lr){5-6}\cmidrule(lr){7-8}
Config & $N$ & off & on & off & on & off & on \\
\midrule
indep. & 1  &  45.5 &  39.6 & 0.65 & 0.68 &  2.70 &  2.72 \\
indep. & 8  & 226.3 & 162.3 & 1.83 & 2.15 & 10.66 & 10.82 \\
indep. & 32 & 348.9 & 199.4 & 5.89 & 7.14 & 27.61 & 27.67 \\
shared & 1  &  46.6 &  41.5 & 0.56 & 0.58 &  2.59 &  2.58 \\
shared & 8  & 420.4 & 267.8 & 1.16 & 1.32 &  5.11 &  5.44 \\
shared & 32 & 474.6 & 240.6 & 3.22 & 3.85 & 10.50 & 10.71 \\
\bottomrule
\end{tabular}
\end{table}

\begin{figure*}[t]
\centering
\includegraphics[width=0.98\textwidth]{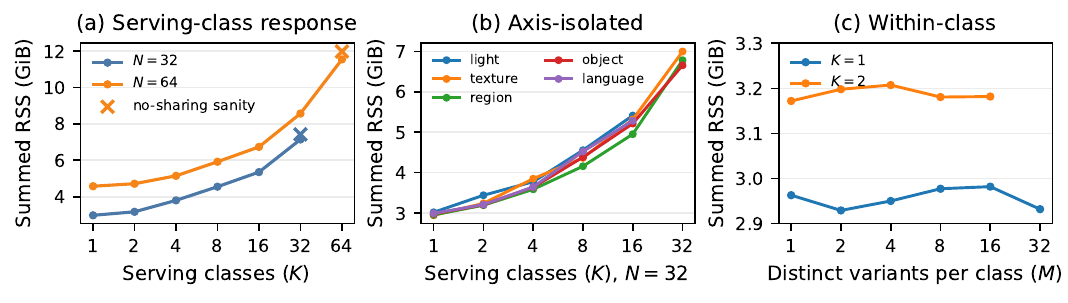}
\caption{LIBERO-plus memory use as template diversity (a, b) and distinct
variants per class (c) increase. Crosses mark independent-model baselines.}
\label{fig:d1-diversity}
\end{figure*}

\paragraph{Camera observations.} We enable one $224{\times}224$ RGB camera
in each environment's step observations (Table~\ref{tab:s2-camera}).
The memory advantage persists: at $N{=}32$, shared-template threads use
3.9 versus 7.1\,GiB of GPU memory and 10.7 versus 27.7\,GiB of summed
RSS for independent threads.
Camera observations reduce shared-template throughput from 475 to 241
steps/s, which remains 21\% above independent threads.
The camera observation path still processes images in Python and is subject
to GIL contention as concurrency grows.
We expect moving this Python observation path into native code to alleviate
the throughput scaling bottleneck.

\subsection{Memory under Workload Diversity}
\label{sec:eval-diversity}

We separate the memory effects of template diversity and unique asset content.
Here $N$ is the number of environment instances and $K$ is the number of
serving classes, each materialized with one template.
These construction experiments do not initialize a renderer.

\paragraph{Controlled variation.} We hold $N$ fixed and increase $K$ in
LIBERO-plus (Figure~\ref{fig:d1-diversity}a).
Each class serves at least one instance, so $K\le N$.
At $N{=}32$, summed RSS grows from 2.98 to 7.14\,GiB as $K$ increases
from 1 to 32.
When $K{=}N$, no two instances reuse a template, and RSS approaches the
independent-model baseline.
The same trend holds at $N{=}64$ and when varying each mutation axis
separately (Figure~\ref{fig:d1-diversity}b).
Light and language sweeps stop at $K{=}16$ because the index contains only
21 eligible classes for those axes.
At fixed $N{=}32$ and $K$, we increase the number $M$ of distinct task variants per class, repeating sampled variants to fill the 32 instances (Figure~\ref{fig:d1-diversity}c).
The constraint $KM\le32$ limits the $K{=}2$ sweep to $M{=}16$.
RSS varies by at most 55\,MiB as the number of distinct variants per class increases from 1 to 32, showing little additional memory cost for task variation served by the same templates.

\begin{figure*}[t]
\centering
\includegraphics[width=0.98\textwidth]{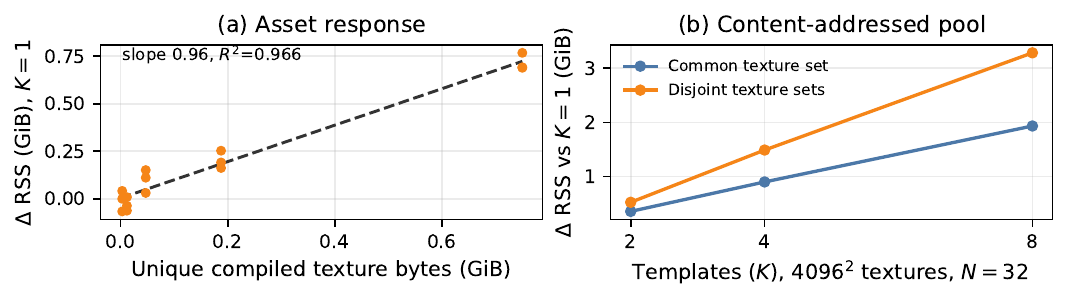}
\caption{RoboCasa365 memory response to unique texture bytes (a) and
texture sharing across templates (b), with $N{=}32$.}
\label{fig:d2-scatter}
\end{figure*}

\paragraph{Controlled texture response.} To isolate asset-content costs,
we generate RoboCasa365 workloads with noise images of controlled
resolutions in the four generative texture slots.
The requested files determine the total unique compiled texture bytes.
With one template, RSS grows approximately linearly with this quantity
over a 0.75\,GiB range, with slope 0.96 and $R^2{=}0.966$
(Figure~\ref{fig:d2-scatter}a).
To test reuse across templates, we compare $K$ templates referencing one
common texture set with $K$ templates referencing disjoint sets
(Figure~\ref{fig:d2-scatter}b).
At $K{=}8$, the disjoint case uses 1.35\,GiB more RSS, consistent with
the cost of seven additional 0.19\,GiB texture sets.
Texture differences still require separate templates under the current
vocabulary, while the content pool reuses textures shared across those
templates.

\paragraph{Native-style workloads.} We also use RoboCasa365's native
texture choices, varying palette overlap across three levels at $K{=}8$
and $N{=}32$ while holding the task, layout, and replay metadata fixed.
Template serving reduces summed RSS by 31--33\%, from 8.41--8.65\,GiB
with independent models to 5.60--5.94\,GiB.
These workloads combine texture overlap with differences in compiled asset
composition.

\subsection{Template Serving and Correctness}
\label{sec:eval-serving}

\begin{table}[t]
\centering
\small
\setlength{\tabcolsep}{5pt}
\caption{Static accounting of the LIBERO-plus offline index.}
\label{tab:index-accounting}
\begin{tabular}{lr}
\toprule
Index quantity & Value \\
\midrule
Variant records & 16{,}170 \\
Distinct models & 5{,}855 \\
Selected templates & 3{,}042 \quad ($-48.0$\%) \\
Additional models served & 2{,}813 \quad (6 zero-write) \\
Patch writes per additional model & median 6, max 22 \\
\bottomrule
\end{tabular}
\end{table}

\paragraph{Offline index.} The 16,170 LIBERO-plus records describe 5,855 distinct models.
Template selection reduces the number of templates needed to serve these models to 3,042, a 48\% reduction (Table~\ref{tab:index-accounting}).
The remaining 2,813 models are served by applying patches to the selected templates.
These patches contain a median of six field writes and a maximum of 22, with six requiring no writes.

% [superseded by fig:selection-policy]
% \begin{table}[t]
% \centering
% \small
% \setlength{\tabcolsep}{4.5pt}
% \caption{Serving policies on uniformly random request draws from the
% LIBERO-plus corpus (medians of three trials).  The offline index resolves
% every request without compilation; incremental caching compiles each
% distinct source once; the baseline compiles every request.}
% \label{tab:selection-policy}
% \begin{tabular}{rrrrr}
% \toprule
%  & \multicolumn{3}{c}{Templates materialized} & \\
% \cmidrule(lr){2-4}
% Draw size & All & Incremental & Offline index & Select.\ (ms/req) \\
% \midrule
% 16  & 16  & 14  & 15  & 0.07 \\
% 64  & 64  & 46  & 45  & 0.26 \\
% 256 & 256 & 144 & 127 & 0.43 \\
% \bottomrule
% \end{tabular}
% \end{table}

\begin{figure}[tb]
\centering
\includegraphics[width=\linewidth]{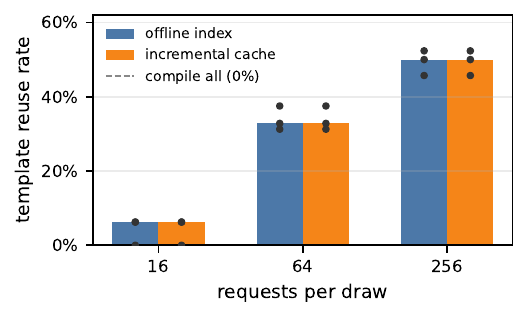}
\caption{Template reuse under uniform LIBERO-plus request sampling.
Bars show medians of three trials, with individual trials marked by dots.}
\label{fig:selection-policy}
\end{figure}

\paragraph{Selection policy.} We compare the offline index and incremental selection on identical uniformly sampled request sets, with three trials at each size (Figure~\ref{fig:selection-policy}).
We measure reuse as $1-T/R$, where $T$ is the number of selected templates and $R$ is the number of requests.
Constructing one model per request gives zero reuse.
Both paths select the same number of templates in every trial, with median reuse rising from 6\% at 16 requests to 50\% at 256 requests.
The offline index preserves this reuse while reducing online assignment cost from 0.79--0.88\,s to 0.51--0.53\,ms per request by eliminating runtime profile construction and patch derivation.
Incremental selection itself takes only 0.04--0.45\,ms once profiles are available, with profile construction accounting for nearly all of its online cost.
Loading the index takes 0.14\,s once per run.
These timings exclude template materialization, which is still required on first use.

\paragraph{Semantic checks.} Model-level acceptance tests compare served profiles with independently compiled target models.
Seeded rollout checks cover four tasks over 30 steps, comparing each member of a shared group with a separately run instance using the same template, patch, seed, and actions.
Simulator states and images from both cameras match bitwise, and success flags agree.
Rebuilt groups and indexed versus incremental selection also produce identical trajectory hashes.

\subsection{Component Costs and Ablation}
\label{sec:eval-components}

\begin{figure}[tb]
\centering
\includegraphics[width=\linewidth]{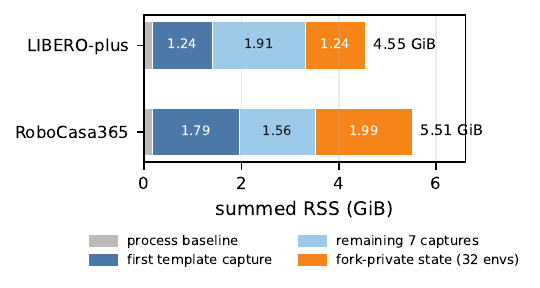}
\caption{RSS growth during template and instance construction, with $K{=}8$ and $N{=}32$.}
\label{fig:layer-attribution}
\end{figure}

% [superseded by fig:layer-attribution]
% \begin{table}[t]
% \centering
% \small
% \setlength{\tabcolsep}{5pt}
% \caption{Physical build profile of the K=8, N=32 workloads: incremental RSS
% sampled after every construction stage (GiB). Texture pixels live in a
% content-addressed pool outside the compiled model, so the stages, not nominal
% array sizes, carry the accounting.}
% \label{tab:layer-attribution}
% \begin{tabular}{lrr}
% \toprule
% Stage & LIBERO & RoboCasa \\
% \midrule
% Process baseline & 0.17 & 0.17 \\
% First template capture & 1.24 & 1.79 \\
% Remaining 7 template captures & 1.91 & 1.56 \\
% Fork-private state (32 envs) & 1.24 & 1.99 \\
% \midrule
% Peak summed RSS & 4.55 & 5.51 \\
% \bottomrule
% \end{tabular}
% \end{table}

\begin{figure*}[t]
\centering
\includegraphics[width=0.98\textwidth]{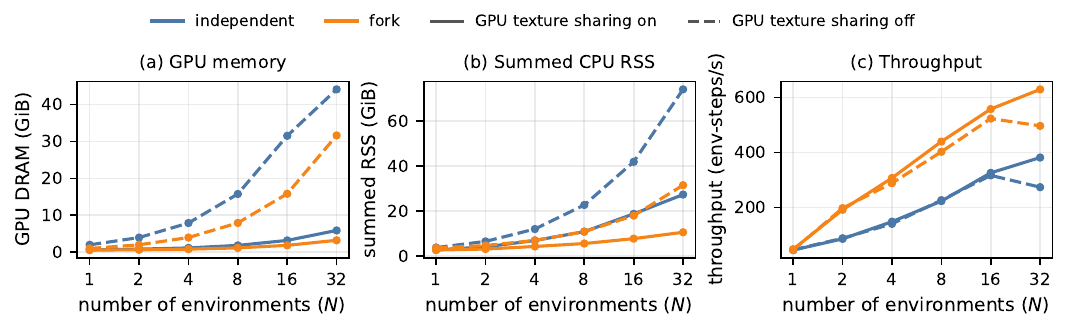}
\caption{Model and GPU texture sharing ablation on RoboCasa365, showing medians of three seeds.}
\label{fig:s1-sharing}
\end{figure*}

\paragraph{Construction costs.} We sample RSS after each template capture and each batch of instances to measure memory growth during construction (Figure~\ref{fig:layer-attribution}).
At $K{=}8$ and $N{=}32$, constructing the first template adds 1.24\,GiB for LIBERO-plus and 1.79\,GiB for RoboCasa365, including initial asset loading and initialization.
The remaining seven templates together add 1.91 and 1.56\,GiB, respectively.
Constructing the 32 instances adds another 1.24 and 1.99\,GiB, averaging about 40 and 64\,MiB per instance.
This per-instance cost includes private environment objects and simulation state.

In the construction microbenchmark, capturing a template from an existing environment takes about 1\,ms, compared with 0.3--0.8\,s to construct and reset the source environment.
Creating an instance from the template takes about 0.1\,s.
Template capture adds little overhead to source construction, and subsequent instances avoid rebuilding the shared model.

% [superseded by fig:s1-sharing, which now carries all three metrics vs N]
% \begin{figure*}[t]
% \centering
% \includegraphics[width=0.78\textwidth]{Sec09-sharing-ablation.pdf}
% \caption{Two-factor ablation at $N{=}32$ (RoboCasa365, state-only, medians
% of three seeds): model-level template sharing (fork) crossed with the
% renderer-side texture registry.  Panel (a) shows GPU and summed-RSS memory
% (log scale); panel (b) shows throughput.  Each
% factor alone reduces memory; combined they cut summed RSS 7.0$\times$ and
% GPU memory 13.7$\times$ against independent models without the registry.}
% \label{fig:sharing-ablation}
% \end{figure*}

% [superseded by fig:sharing-ablation]
% \begin{table}[t]
% \centering
% \small
% \setlength{\tabcolsep}{5pt}
% \caption{Two-factor ablation at $N{=}32$ (RoboCasa365, state-only, medians
% of three seeds): model-level template sharing (fork) crossed with the
% renderer-side texture registry.  GPU memory and summed RSS in GiB.  Each
% factor alone reduces memory; combined they cut summed RSS 7.0$\times$ and
% GPU memory 13.7$\times$ against independent models without the registry.}
% \label{tab:sharing-ablation}
% \begin{tabular}{llrrr}
% \toprule
% Models & Registry & GPU & RSS & Steps/s \\
% \midrule
% independent & off & 44.12 & 73.99 & 273.6 \\
% independent & on  &  5.89 & 27.24 & 381.5 \\
% fork        & off & 31.63 & 31.42 & 496.7 \\
% fork        & on  &  3.22 & 10.49 & 629.6 \\
% \bottomrule
% \end{tabular}
% \end{table}

\paragraph{Sharing-layer ablation.} Figure~\ref{fig:s1-sharing} independently enables model sharing and GPU texture sharing across replica counts.
At $N{=}32$, model sharing alone lowers summed RSS from 74.0 to 31.4\,GiB by reusing the captured model and its host-side content.
GPU texture sharing alone lowers GPU memory from 44.1 to 5.9\,GiB by reusing textures across rendering contexts.
Enabling both lowers summed RSS to 10.5\,GiB and GPU memory to 3.2\,GiB, reductions of 7.0$\times$ and 13.7$\times$ from the configuration with both disabled.
The combined savings show that sharing at both layers is needed to reduce CPU and GPU memory together.
Throughput also improves with either layer enabled, reaching 629.6 steps/s with both enabled compared with 273.6 steps/s with both disabled.

\section{Discussion}
\label{sec:discussion}

\paragraph{Independent Agent Interactions and Batched Simulation.}

Simulation sandbox workloads can differ from lockstep reinforcement-learning rollouts.
An agent may need to keep interacting with one environment while other agents are waiting for model inference, diagnosis, or recovery decisions.
GPU vectorized simulators are highly effective when environments advance in synchronized batches, but irregular arrivals can leave a runtime choosing between waiting to form a batch and executing smaller batches.
We focus on CPU-based simulation sandboxes, where environments can progress independently while still sharing resources within a process.
This execution flexibility motivates our target workload; quantifying the trade-off against GPU batching under asynchronous agent requests remains future work.

\paragraph{Sharing Boundaries and Generality.}
The sharing opportunities a runtime can realize depend on its resource representation and supported mutations.
A field's size and the updates it induces in derived state help determine whether it can vary privately while the rest of a model remains shared.
Our current backend restricts these mutations to keep their effects and isolation requirements explicit.
Equivalence-based serving then provides a concrete policy within the broader serving preorder.
A richer backend could support asset rebinding and union templates that serve targets from several current equivalence classes, at the cost of additional storage, isolation, and validation requirements.
The template--patch interface allows these policies to evolve alongside the sharing mechanisms, while retaining the requirement that each served instance reproduce its target environment.

\section{Related Work}
\label{sec:related-works}

\paragraph{Sharing in Heterogeneous Simulation.}

% Isaac Lab facts checked against local revision a8b4da3c29ae528b39d4b3c9444d782ce58d886d, 2026-10-18.
GPU simulators combine resource sharing with several forms of environment variation.
MuJoCo Warp supports both shared and per-world model fields; its per-world mesh workflow uses a base model containing all candidate meshes and the maximum required geometry slots, with per-world fields selecting meshes and disabling unused slots~\cite{mujoco-warp-docs}.
Isaac Lab combines USD instancing with backend-native physics replication, mapping asset prototypes to selected environments so that compatible heterogeneous variants can retain physics replication~\cite{isaaclab-cloning-docs}.
Some forms of variation require changing the sharing arrangement: its visual texture randomization implementation disables physics replication and removes instancing from affected visual prims before editing their materials~\cite{isaaclab-visual-randomization}.

\paragraph{Resource Sharing in Sandboxes and Serverless Systems.}
Memory sharing has long supported dense deployment of isolated instances.
Page deduplication and shared file-backed mappings reduce repeated memory across virtual machines and sandboxes, as in VMware ESX, Linux KSM, and DeepSeek's DSec~\cite{waldspurger02,ksm,dsec}.
Serverless systems reuse initialized state through zygote processes, checkpoint restoration, and local or remote fork, reducing startup costs and repeated initialization across instances~\cite{sock,catalyzer,mitosis}.
Faasm combines WebAssembly-based isolation with shared memory regions between co-located functions~\cite{faasm}.
These mechanisms provide reuse at page, process, or sandbox boundaries.
\sys instead derives sharing boundaries from simulation resources and supported mutations, allowing common assets to be reused across different model templates while applying parameter changes to private storage.

\paragraph{Structure-Aware Resource Sharing.}
Application structure can expose reuse opportunities and guide the mechanisms that realize them.
Tetris identifies identical model tensors across serverless functions and maps them into shared read-only memory~\cite{tetris}.
PagedAttention manages KV cache in shareable blocks~\cite{pagedattention}, while SGLang discovers common prefixes through a radix tree and schedules requests to reuse cached state~\cite{sglang}.
Parrot exposes application dataflow through Semantic Variables, enabling the serving system to identify related requests and coordinate their execution~\cite{parrot}.
Sharing also affects memory traffic: FlashInfer's cascade attention separates shared-prefix and private-suffix attention so that multiple queries can reuse prefix KV data within a kernel~\cite{flashinfer,cascade-inference}.
Across these systems, workload representations, reuse discovery, and execution mechanisms work together to exploit common structure.
\sys follows this approach for simulation: explicit mutations and resource profiles expose reuse, serving policies select templates, and the runtime materializes the assignments through model and asset sharing.

\section{Conclusion}
\label{sec:conclusion}

\sys represents domain-randomized environments as shared templates and explicit patches, allowing simulation sandboxes to reuse common resources across task variants.
Its mutation interface, layered sharing runtime, and template-serving policies connect environment differences to the physical resources that must be private or can remain shared.
Our robosuite/MuJoCo prototype reduces memory use in replica-scaling and controlled heterogeneous workloads, with savings determined by template reuse and asset overlap.
These results support template-based resource sharing as a way to deploy diverse simulation environments within a fixed memory budget.

%-------------------------------------------------------------------------------
\bibliographystyle{plain}
\bibliography{refs}

%%%%%%%%%%%%%%%%%%%%%%%%%%%%%%%%%%%%%%%%%%%%%%%%%%%%%%%%%%%%%%%%%%%%%%%%%%%%%%%%
\end{document}